\documentclass[]{spie}  
\usepackage[]{graphicx}
\usepackage{url}
\usepackage{epstopdf}  
\usepackage[labelformat=empty]{caption}
\usepackage[table]{xcolor}
\usepackage{subcaption}
\usepackage{ams math} 
\usepackage[export]{adjustbox}
\usepackage{color}
\usepackage[load=prefix,load=named]{siunitx}
\usepackage{multirow}

\title{Deploying quantum light sources on nanosatellites II: lessons and perspectives on CubeSat spacecraft} 

\author{R. Bedington\supit{a}, E. Truong-Cao\supit{a}, Tan Y. C.\supit{a}, C. Cheng\supit{a}, K. Durak\supit{a}, J. Grieve\supit{a}, J. Larsen\supit{d}, D. Oi\supit{b} and A. Ling\supit{a,c}
\skiplinehalf
\supit{a}Centre for Quantum Technologies, Block S15, 3 Science Drive 2, National University of Singapore, 117543, Singapore \\
\supit{b} SUPA Dept of Physics, University of Strathclyde, 107 Rottenrow East, Glasgow G4 0NG, United Kingdom\\
\supit{c} Dept of Physics, National University of Singapore, 2 Science Drive 3, Singapore, 117542, Singapore\\
\supit{d} Dept of Electronic Systems, Aalborg University, Fredrik Bajers Vej 7, 9220 Aalborg, DK
}

\authorinfo{Further author information: Send correspondence to A. Ling, E-mail: cqtalej@nus.edu.sg, or R. Bedington, E-mail: r.bedington@nus.edu.sg }

\begin{document} 
  \maketitle 

\begin{abstract}
To enable space-based quantum key distribution proposals the Centre for Quantum Technologies is developing a source of entangled photons ruggedized to survive deployment in space and greatly miniaturised so that it conforms to the strict form factor and power requirements of a 1U CubeSat. The Small Photon Entangling Quantum System is an integrated instrument where the pump, photon pair source and detectors are combined within a single optical tray and electronics package that is no larger than 10 cm x 10 cm x 3 cm. This footprint enables the instrument to be placed onboard nanosatellites or the CubeLab structure aboard the International Space Station. We will discuss the challenges and future prospects of CubeSat-based missions.
\end{abstract}



\section{INTRODUCTION}
\label{sec:intro}  

Quantum Key Distrbution (QKD) comprises a family of cryptographic schemes motivated by the enhanced privacy guarantees from quantum mechanics. One of the challenges in QKD research is to extend coverage to a global network. Single photons cannot be re-amplified without destroying their quantum properties. This limits practical fiber-based QKD to about \SI{100}{\km} after which trusted nodes (which rely on traditional security mechanisms) or quantum repeaters (which are not yet demonstrated) will be necessary. 

Terrestrial free-space QKD is limited by the need for line-of-sight locations with the current record at \SI{150}{\km}. Greater distance can be achieved  by increasing the altitude of the quantum transceivers, e.g. by using long duration high altitude platforms. A compelling pathway to global QKD networks is to place core pieces of enabling technology on satellites in low Earth orbit (LEO). A number of proposals have been been published for building global quantum communication networks using satellites that host quantum light sources or detectors \cite{ursin09,ling12,morong12,scheidl13,jennewein14}. Efforts are under way to implement the first demonstrations. 

\begin{figure}[!h]
\centering
\begin{minipage}[t]{0.8\textwidth}
\centering
\includegraphics[width=0.8\textwidth,keepaspectratio]{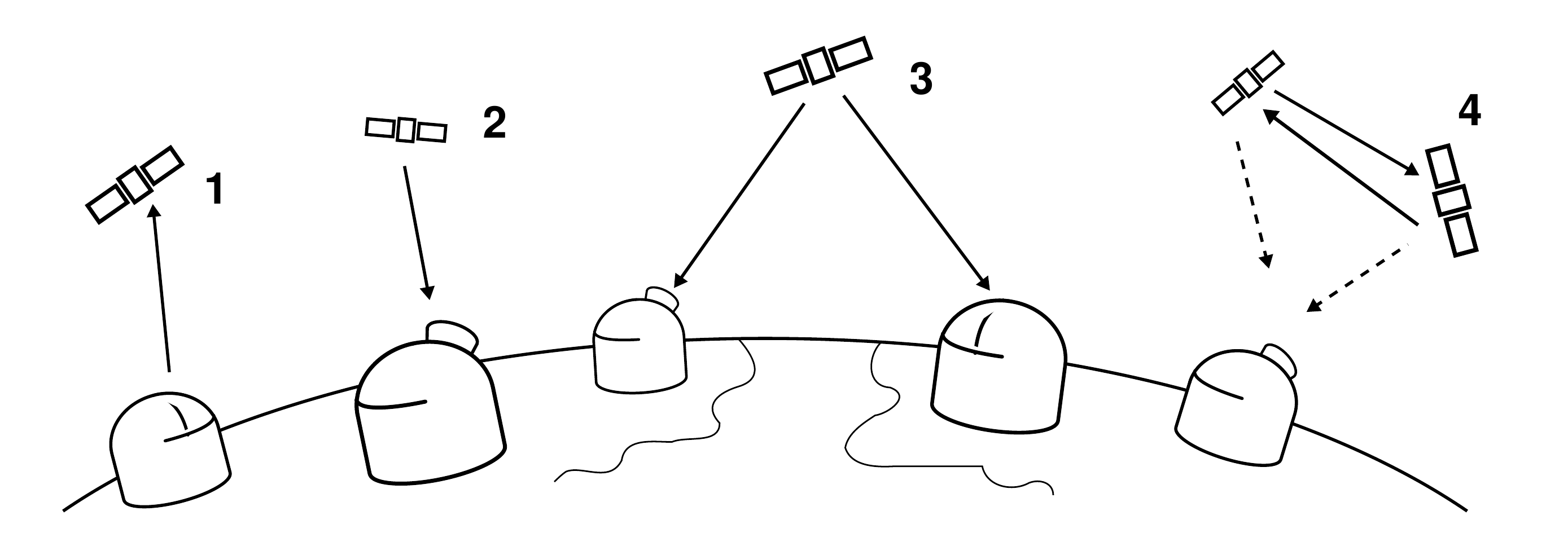}
\vspace{8pt}
\caption{{\small \textbf{Figure 1.} Possible satellite-based QKD type experiments. 1. Use of an uplink where the satellite only carries single photon detectors. 2. A downlink configuration where the satellite carries a source and detectors. 3. A platform that can beam to two ground stations simultaneously (this was the original Space-QUEST concept). 4. Inter-satellite QKD which could be the building block for a long baseline test of quantum correlations. To enable configurations 2-4 with Bell-type measurements, a source of entangled photons in space must be demonstrated. }}
\label{fig:schemes-2}
\end{minipage}
\end{figure}

In the scenario that we envision a source of strongly correlated photons will be placed aboard cost-effective satellites and the photons will be beamed to receiving ground stations \cite{ling12}. An alternative satellite-based approach is to put only single photon receivers on the satellite \cite{jennewein14}. This detector-based approach, however, suffers from additional link loss \cite{bourgoin13} and does not pave the way for inter-satellite links which may be of interest when it comes to long baseline tests of quantum correlations (see Fig. \ref{fig:schemes-2}).

In the simplest implementation of space-based QKD the satellite and a ground station will establish a secret key. In a more advanced implementation the satellite will operate as a trusted node between multiple ground stations. Although trusted nodes rely on non-quantum security assurances a space-based node is difficult to access via side-channels. For example the reported side-channel attacks on terrestrial QKD systems \cite{gerhardt11} are difficult to implement when targeting a fast-flying node where communications can only be achieved over a solid angle of tens of pico steradians. While the pointing accuracy is a technical challenge it is worthwhile to note that optical links between satellites and ground stations already exist \cite{opals14}. The missing key enabling technology in this scenario is a working space-capable source of quantum correlated photons.

We have proposed that nanosatellites (spacecraft that have a mass below \SI{50}{\kg}) have a role to play in this effort \cite{ling12,morong12}. They could act as demonstrators to raise the technology readiness level (TRL) of essential components and also as the final platforms that transmit and receive single photons from ground-based stations or other satellites. In particular nanosatellites can effectively host robust and compact sources of polarization-entangled photon pairs which are the workhorse for entanglement-based QKD. The decreasing cost of launching a nanosatellite into low Earth orbit \cite{coopersmith11} has added impetus to this approach.

\begin{figure}[!h]
\centering
\begin{minipage}[t]{0.8\textwidth}
\centering
\includegraphics[width=0.8\textwidth,keepaspectratio]{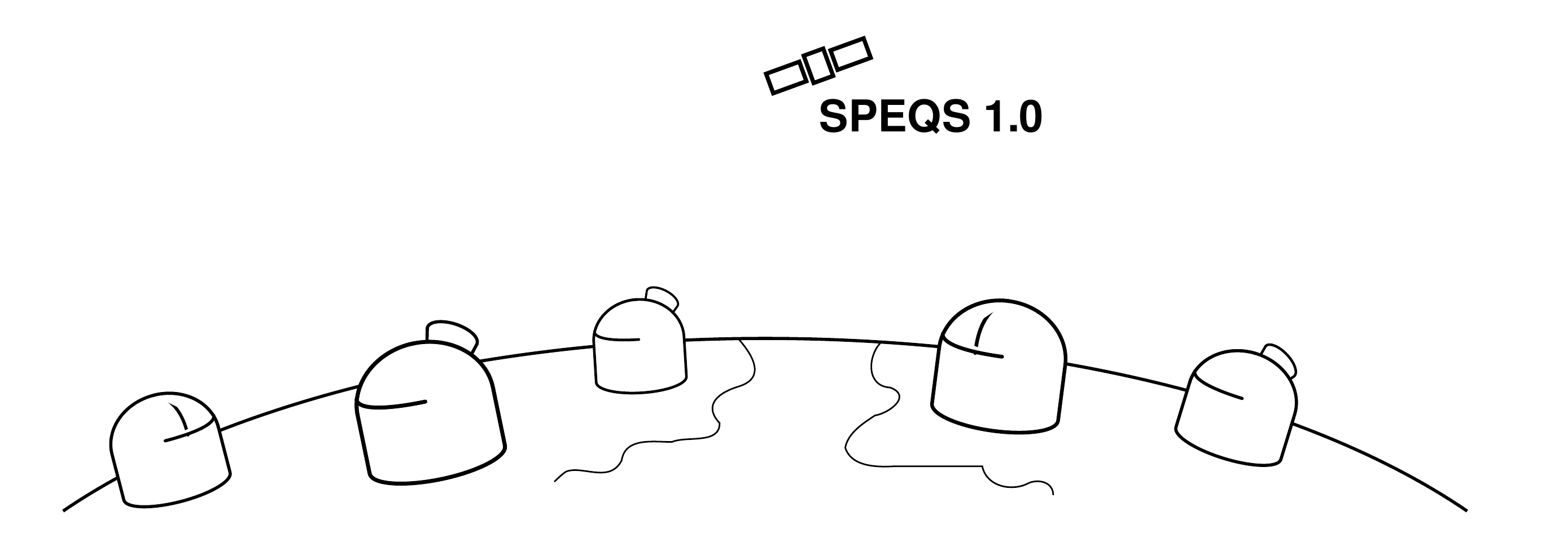}
\vspace{8pt}
\caption{{\small \textbf{Figure 2.} In our first satellite mission we aim to deploy a small and rugged correlated photon source on a (nano-)satellite to verify the ruggedness of the optical design and the operation of the supporting electronics. No optical link is needed in this configuration. }}
\label{fig:schemes-2-cqt}
\end{minipage}
\end{figure}

Our approach to this task is to take iterative steps towards a final demonstration of entanglement-based QKD from space platforms \cite{ling12}. The immediate task is to demonstrate that the basic optical design is rugged and that the control electronics are able to operate within the expected environmental envelope in LEO. To increase the chances of such a demonstration we have designed the instruments to be compatible with a widely used nanosatellite CubeSat standard \cite{woellert11}. We have designed a basic demonstration unit that minimizes size, weight and power (SWAP) requirements to increase the chance of acceptance onto a spacecraft. For this first step pointing by the host spacecraft is not needed (see Fig. \ref{fig:schemes-2-cqt}).

The demonstration photon pair source that we are building is called the Small Photon-Entangling Quantum System 1.0 (SPEQS-1.0). It is an integrated instrument combining low-power electronics and a rugged optical assembly. The goal of the SPEQS-1.0 instrument is to demonstrate that an entangled photon source works in LEO. This device has been extensively tested and documented \cite{tan13,tang_icsos_14,tang14,tan15,chandrasekara15_spie,cheng15,chandrasekara15_spie2} and has undergone two launch campaigns \cite{gomx2,luo14}. It is has been assessed to be at a TRL of 8 \cite{trl}.
In this article we will focus on the advantages and prospects of working with CubeSats and describe the initial plans for a dedicated CubeSat to support the next-generation SPEQS device. We will conclude with a discussion on how and if the CubeSat concept can truly be used to carry out a full space-to-ground QKD demonstration.

\section{CubeSats and SWAP restrictions} \label{sec:cubesat}
The primary building block of a free-flying CubeSat spacecraft is a \SI{10}{cm} cube that is a fully-functional satellite containing the main components expected within a satellite bus (e.g. solar panels, batteries, onboard computer and radio transceiver). The CubeSat architecture allows the basic cube (1U) to be stacked into larger sized spacecraft composed of multiple cubes with the 2U and 3U sizes being currently quite popular (fractional sizes are not so common). In the future it is expected that CubeSat-based spacecraft would reach 6U or larger as these will have sufficient resources to achieve more complex mission requirements.

CubeSat based spacecraft have become very popular for amateur researchers, university research groups and technology companies seeking a market niche. CubeSats are typically built from commercial-off-the-shelf components (COTS) as these satellites are usually launched into orbits with an orbital life that is measured in years or even months \cite{swartwout14} (although some CubeSats have been operational now for 7 years \cite{bouwmeester08}). Space heritage of components is acquired from the increasing number of successful launches that share common parts. The approach of using COTS components has been highly beneficial to the SPEQS program as it has enabled the developers to focus on the design of the instrument using mature technologies that have not been formally rated for space. This avoids the use of expensive and scarce space-rated components that may have been designed to greatly outperform the requirements of a CubeSat-based mission. Space readiness is then acquired by carrying out targeted testing. This strategy has helped to reduce the necessary development time and to introduce savings enabling our small team to prepare for a space mission.

CubeSats typically get launched into orbit by piggy-backing on larger satellite launches. A number of launch brokers \cite{nanoracks,spaceflight} help CubeSat developers get launched from traditional launchers or via the International Space Station. Piggy-backing helps make launch costs accessible to university groups. But there is a further level of savings available where each CubeSat is used to host multiple research payloads. A drawback for this approach is the low duty-cycle for experiment time as the limited power on a CubeSat results in restricted operation time for each payload.

A limitation of current CubeSat technology is that it fails to meet the requirements of high-performance optical transmission experiments as there has been no demonstration of fine pointing on the order of a few micro-radians which is necessary for space-to-ground optical communications (crude optical communication based on optical Morse code has been demonstrated \cite{fitsat1}). Our team is in discussion with collaborators about the feasibility of such a spacecraft and we are aware that since we have proposed the CubeSat approach several other groups have recently expressed interest in this technology \cite{jennewein14}. 

\begin{figure}[!h]
\centering
\begin{minipage}[t]{0.8\textwidth}
\centering
\includegraphics[width=1.0\textwidth,keepaspectratio]{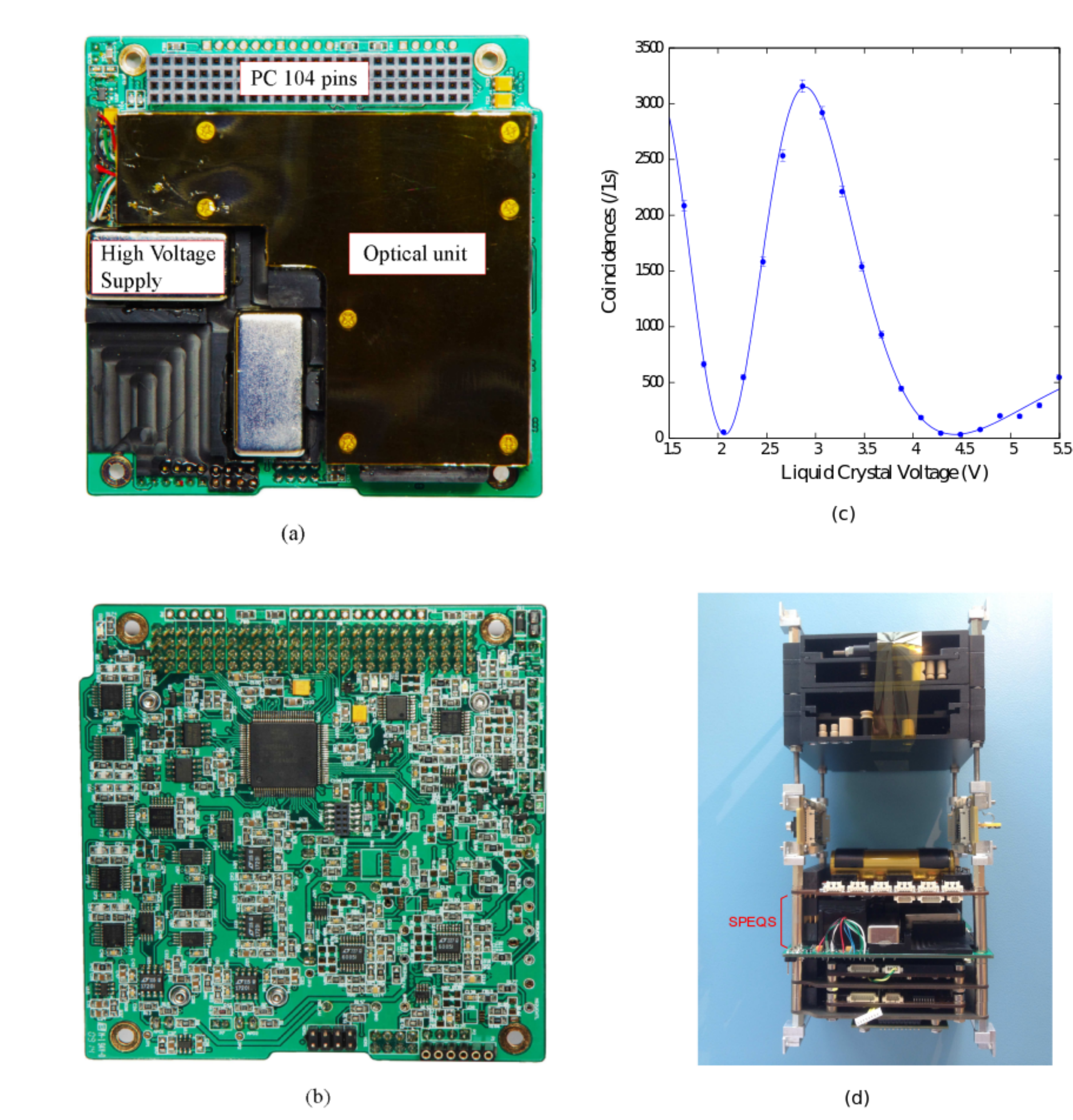}
\vspace{8pt}
\caption{{\small \textbf{Figure 3.} (a) The SPEQS-1.0 optical unit is contained within a light-tight aluminium 7075 tray and mounted to the printed circuit board. (b) The SPEQS-1.0 electronics platform on the reverse side of the printed circuit board. This board also provides the mechanical interface to the spacecraft using the four corner screw slots and the PC104 bus that also provides power and intra-satellite communications. (c) Typical polarization correlations reported by the instrument. (d) The SPEQS-1.0 instrument integrated with the GomX-2 spacecraft. Published with permission from GomSpace ApS.}}
\label{fig:speqs}
\end{minipage}
\end{figure}

Entangled photon sources for nanosatellite operations must satisfy the SWAP criteria while having the necessary brightness, entanglement quality and the potential for ruggedization. 
The SWAP restrictions for CubeSats are very stringent. For CubeSats hosting multiple payloads the most critical limit is size. An entire entangled photon source containing also Bell-state analyzer apparatus must be contained in a volume that is approximately \SI{10}{\cm}$\times$\SI{10}{\cm}$\times$\SI{3}{\cm}. The overall mass of such an instrument should not exceed \SI{500}{\g}, while the continuous (peak) power consumption should be below \SI{2.0}{\W} (\SI{2.5}{\W}).

These restrictions lead to a few defining requirements for the source. First non-collinear SPDC sources are ruled out because they require more complex (and less robust) arrangements to achieve a compatible form factor. After weighing the various aspects between different source designs it was decided to proceed with a geometry based on $\beta$-Barium Borate  crystals \cite{trojek08,tang14}. The current form-factor for the SPEQS-1.0 instrument and its typical polarization correlation output is illustrated in Fig. \ref{fig:speqs}. The internal geometry is described elsewhere \cite{tang14}. The instrument has been successfully integrated and qualified for flight with a CubeSat (see Fig. \ref{fig:speqs}).

A final hurdle for CubeSat missions is attempting to get into an appropriate orbit in a reasonable time. Anecdotal evidence suggests that while launching as piggy-backs offers many opportunities the reality is slightly more subtle with numerous delays being encountered. For this reason launching via the International Space Station (ISS) is becoming increasingly popular although this is not without risk. In the last year three launch vehicles to the ISS were lost resulting in dozens of nanosatellites being destroyed. The instrument shown in Fig. \ref{fig:speqs} was lost when a launch vehicle to the ISS failed. Despite the relative low cost of CubeSats it is often difficult for university teams to prepare spare spacecraft for re-launch even if these follow-up opportunities come at much reduced cost. For a robust program it is prudent to ensure that a spare flyer is available. 

\section{SpooQy-Sat} \label{sec:spooqy}
\subsection{Proposed development plan} \label{sec:spooqyplan}
In the next generation SPEQS-2.0 device one goal is to develop a source that can demonstrate at least one million coincidence pairs a second divided equally between two polarization bases \cite{chandrasekara15_spie2}. This will require optical components such as optics for the pump and collection beams and necessitates a form-factor that already exceeds standard CubeSat payloads. This changes to form-factor prevent us from following the cost-sharing model where we deploy the instrument on multi-mission CubeSats. This has required the development of a dedicated satellite. We have begun a program to develop a series of CubeSats dedicated to demonstrating the SPEQS-2.0 devices. These series of satellites are known as the SpooQy-Sats. 
\begin{figure}[!h]
\begin{center}
 \includegraphics[scale=0.3]{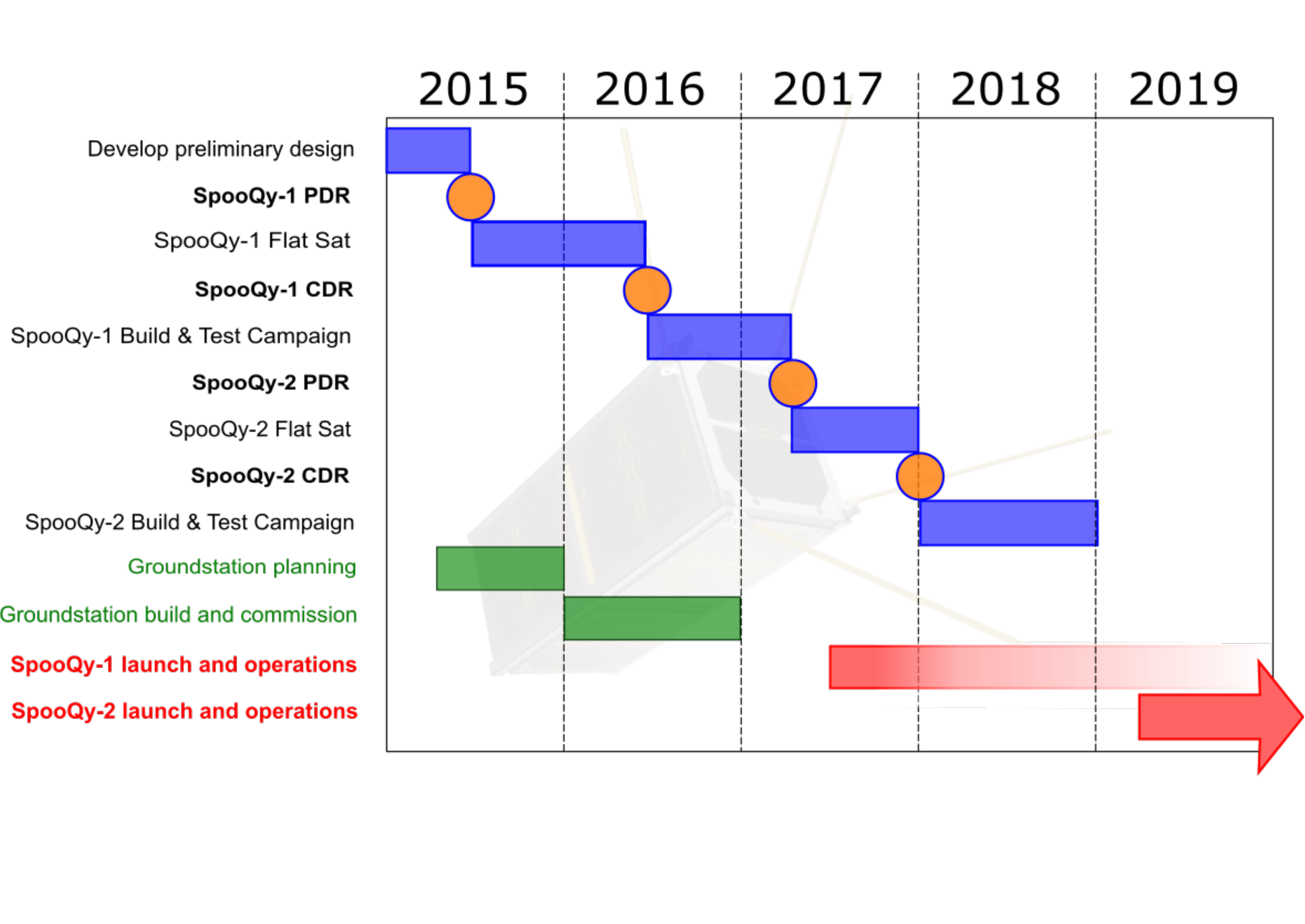}
\caption{\small {\textbf Figure 4.} The proposed SpooQy-Sat development schedule. Included is the proposed development plan for the supporting ground station and SPEQS-2.0.}
\label{fig:spooqysched}
\end{center}
\end{figure}

Two SpooQy-Sats (SpooQy-1 and SpooQy-2) are being planned. The proposed schedule for SpooQy-Sat development is shown in Fig. \ref{fig:spooqysched}. As CubeSats are a novel technology tool for quantum optics research teams the first iteration is planned as an educational exercise. This first spacecraft (SpooQy-1) and has undergone a preliminary design review (PDR). As pointed out by Swartwout \cite{swartwout14} a large number of university based CubeSats fail very rapidly in orbit due to both lack of experience and lack of system-level testing. To de-risk the SpooQy-Sat series veteran satellite developers have been recruited onto the review panel to add depth and experience into system, risk and mission analysis. A range of environmental tests are also being planned including stress tests within a thermal-vacuum chamber. The SPEQS-2.0 payload for both spacecraft will be subject to its own design review due in 2016.

\begin{figure}[!h]
\begin{center}
 \includegraphics[scale=0.2]{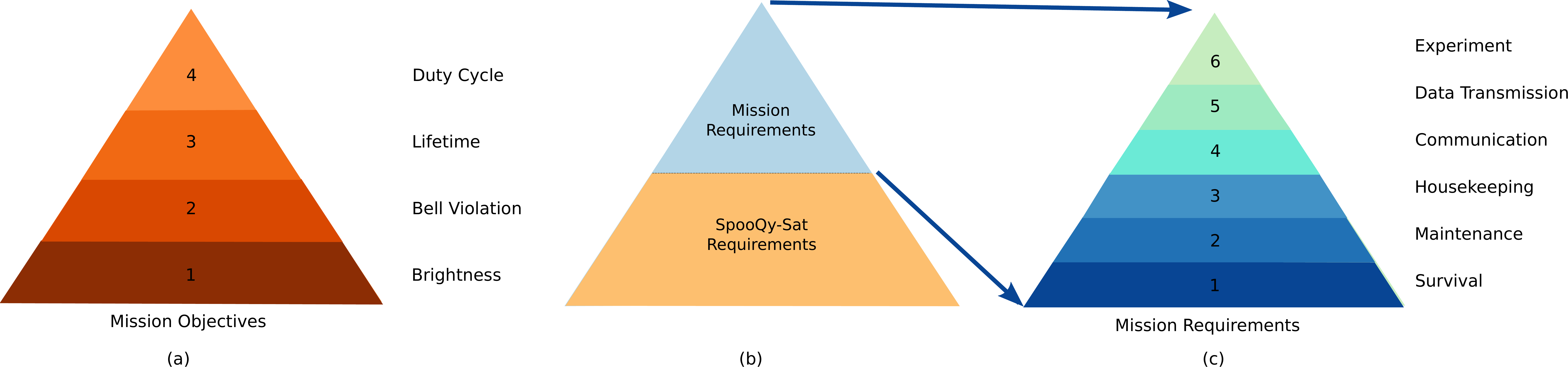}
\caption{\small {\textbf Figure 5.} Proposed hierachy of objectives and requirements for the SpooQy-Sat missions. (a) Four mission objectives are planned. The minimum level of success is the demonstration of one million detected pairs a second distributed over two polarization bases. (b) Spacecraft requirements are in the process of being prepared. These have to support the testable mission requirements defined in (c).}
\label{fig:objectives}
\end{center}
\end{figure}

The mission objectives are illustrated in Fig. \ref{fig:objectives}(a). The minimum level of success is to illustrate that the new optical payload can generate and detect one million pairs of photons per second. The second objective is achieved when it is illustrated that these photons can violate a Bell Inequality. The third objective would be to demonstrate that the photon pairs can operate for a minimum lifetime currently estimated at 3 months. This lifetime is determined by the expected spacecraft life cycle when launched from the ISS \cite{qiao13}. The final objective is to collect performance data on the source as it ages keeping track of duty cycle and experiment runs.

A series of traceable and verifiable requirements (Fig. \ref{fig:objectives} (b) and (c)) are necessary to design a spacecraft that can best support the defined mission objectives. The SpooQy-Sat requirements are currently being defined to support the top-level mission requirements. These top-level requirements are divided into six parts. The first requirement is to demonstrate that the integrated spacecraft survives the defined environmental tests. The second requirement is to demonstrate that the spacecraft can provide the necessary power at all times to maintain the SPEQS-2.0 instrument within a temperature range currently expected to be between \SI{20}{\celsius} and \SI{25}{\celsius}. This maintenance mode is necessary to prevent premature aging of the instrument and is expected even in an actual space QKD experiment. The next requirement is that the payload must be able to collect and store housekeeping data. The fourth and fifth requirements are to demonstrate that the payload can communicate with the satellite bus and that all payload data can be transmitted to ground stations. The final requirement is to demonstrate that the spacecraft can meet the payload resource envelope after deployment in orbit.

\subsection{Proposed platform and SWAP envelope} \label{sec:spooqyplat}

\begin{figure}[!h]
\begin{center}
 \includegraphics[scale=0.3]{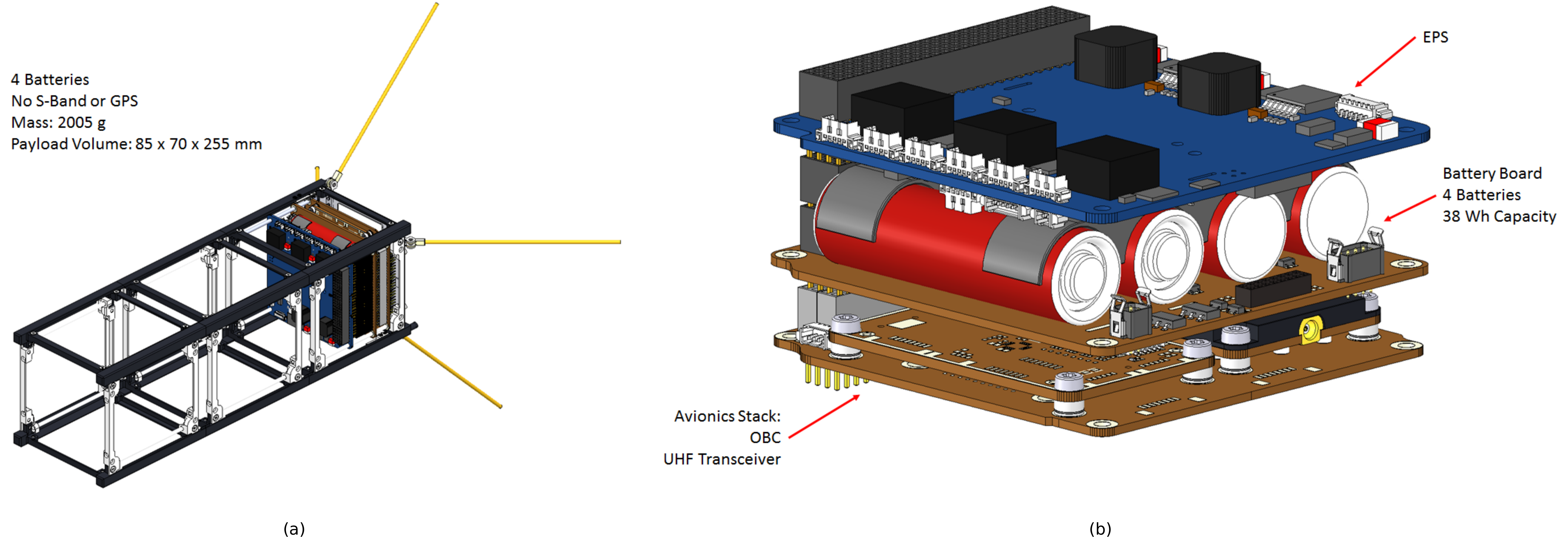}
\caption{\small {\textbf Figure 6.} (a) Proposed SpooQy-1 structure showing satellite bus and reference board with expected volume for SPEQS-2.0 payloads. This structure carries 4 batteries. (b) Preliminary view of the satellite avionics stack based on available off-the-shelf components. }
\label{fig:spoodqysat}
\end{center}
\end{figure}

Prior to the design process a development path survey was conducted to identify suitable CubeSat platforms. The current SpooQy-Sat design uses the GomX platform \cite{gomspace} and the model for a complete SpooQy-Sat bus and structure is shown in Fig. \ref{fig:spoodqysat}. The proposed platform currently has limited flight heritage but is due to be flown on GomX-3 in late 2015 and from other flight opportunities in 2016. The SpooQy-Sat development team is working closely with the GomSpace developers to verify the performance of the proposed platform.

When the proposed spacecraft without payloads is fully assembled with four batteries (38 Wh capacity) it will have an estimated mass of \SI{2.0}{\kg}.  These values are well within the mass budget provided by the Nanoracks interface control document \cite{nanoracksicd} which is \SI{8.4}{\kg}, leaving almost \SI{6.4}{\kg} for the payload volume. Other launch providers have similar mass limits. This mass envelope and the payload volume of \SI{8.5}{\cm}$\times$\SI{7.5}{\cm}$\times$\SI{25.5}{\cm} should accomodate two copies of the proposed SPEQS-2.0 instrument. The estimated average power generated per orbit is \SI{5.8}{\W} and in sunlight instantaneous power is approximately \SI{9}{\W} - it is proposed that the payload operates only during daylight to reduce stress on the electrical power system.

\subsection{Orbit consideration and data links} \label{sec:spooqyorbit}
The choice of orbit determines the overflight duration over the ground station placing a limit on the possible amount of data downloadable for a given hardware specification. Furthermore the amount of solar illumination could also limit the power generation from the solar panels. Our current primary orbit of interest is the ISS orbit for its regular launch schedule and relatively lower launch cost. For comparison we also look at an orbit of approximately 700 km for two different inclinations (equatorial at $20^{\circ}$ and polar sun-synchronous at $98^{\circ}$). 

\begin{table}[h]
\caption{Table 1. Consequences of different orbits on SpooQy-Sat parameters. The average number of passes per day and the average access time per pass is obtained assuming a Singapore based ground station for two different field-of-view values: $140^{\circ}$ and $180^{\circ}$ (unobstructed). Data is obtained from STK simulation.}
\begin{center}       
\begin{tabular}{|c|c|c|c|c|c|} 
\hline
\rule[-1ex]{0pt}{3.5ex}  
Altitude & Inclination & Field-of-view & Average number & Average access & Average sunlight \\
(km) & (deg) & (deg) & of daily passes & time per pass(s) & time per orbit (s) \\ 
\hline
\multirow{2}{*}{400} & \multirow{2}{*}{52} & 140 & 1.7 & 211 & \multirow{2}{*}{3442}  \\
\cline{3-5}
& & 180 & 4.3 & 514 & \\
\hline 
\multirow{2}{*}{700} & \multirow{2}{*}{20} & 140 & 5 & 211 & \multirow{2}{*}{3773}  \\
\cline{3-5}
& & 180 & 12.6 & 514 & \\ 
\hline
\multirow{2}{*}{700} & \multirow{2}{*}{98} & 140 & 1.9 & 310 & \multirow{2}{*}{3910}  \\
\cline{3-5}
& & 180 & 3.9 & 673 & \\ 
\hline
\end{tabular}
\end{center}
\end{table} 

From Table 1 the average sunlight time for different orbits are quite similar. Not shown is the data from a possible dawn-dusk sun synchronous orbit (DD-SSO). Although the DD-SSO is good for electrical power generation the constant sunlight illumination can heat up the satellite and create a temperature gradient across the satellite which is undesirable for the opto-electronics and the optical alignment for the science mission. For the communication system UHF at 9.6kbps and 19.2kbs is being considered. S-band radio was considered initially but it was dropped given the price and maturity of currently available solutions. Table 2 shows the average maximum downloadable data per day for several possible orbits.

\begin{table}[h]
\caption{Table 2. Consequences of different orbits on SpooQy-Sat downloadable data. These values are based on the average number of passes and access time from Table 1.}
\begin{center}       
\begin{tabular}{|c|c|c|c|c|} 
\hline
\rule[-1ex]{0pt}{3.5ex}  
Altitude & Inclination & Field-of-view & \multicolumn{2}{c|}{Average data rate per day (kB)} \\ \cline{4-5}
(km) & (deg) & (deg) & UHF 9.6 kbps & UHF 19.2 kbps \\ 
\hline
\multirow{2}{*}{400} & \multirow{2}{*}{52} & 140 & 423.5 & 847.1  \\
\cline{3-5}
& & 180 & 2581 & 5162  \\
\hline 
\multirow{2}{*}{700} & \multirow{2}{*}{20} & 140 & 2045 & 4090 \\
\cline{3-5}
& & 180 & 22300 & 116000 \\ 
\hline
\multirow{2}{*}{700} & \multirow{2}{*}{98} & 140 & 674 & 1349  \\
\cline{3-5}
& & 180 & 3044 & 6088 \\ 
\hline
\end{tabular}
\end{center}
\end{table} 

The data to be downloaded is dominated by the experiment and is estimated to be 128kB per experiment cycle based on SPEQS-1 experiments. Currently the power budget should enable 12 cycles per day but not all orbits allow sufficient access time for the data to be transmitted to a single ground station in a single 24 hour period. The following table shows the expected number of experiment runs that could be carried out capped to the maximum of 12 cycles per day.

\begin{table}[h]
\caption{Table 3. Consequences of different orbits on the number of accessible SpooQy-Sat experiments. These values are based on the average number of passes and access time from Table 1.}
\begin{center}       
\begin{tabular}{|c|c|c|c|c|} 
\hline
\rule[-1ex]{0pt}{3.5ex}  
Altitude & Inclination & Field-of-view & \multicolumn{2}{c|}{No. of experiment runs per day (max of 12 runs)} \\ \cline{4-5}
(km) & (deg) & (deg) & UHF 9.6 kbps & UHF 19.2 kbps \\ 
\hline
\multirow{2}{*}{400} & \multirow{2}{*}{52} & 140 & 3.3 & 6.6  \\
\cline{3-5}
& & 180 & 12 & 12  \\
\hline 
\multirow{2}{*}{700} & \multirow{2}{*}{20} & 140 & 12 & 12 \\
\cline{3-5}
& & 180 & 12 & 12 \\ 
\hline
\multirow{2}{*}{700} & \multirow{2}{*}{98} & 140 & 5.3 & 10.5  \\
\cline{3-5}
& & 180 & 12 & 12 \\ 
\hline
\end{tabular}
\end{center}
\end{table} 

Based on the results presented in Tables 1-3 the SpooQySat development team is planning to establish a single UHF-based ground station using amateur radio frequencies. A second ground station is desirable for extended reliability and the feasibility of a second ground station in Europe is currently being studied. 

\begin{figure}[!h]
\begin{center}
\includegraphics[scale=0.45]{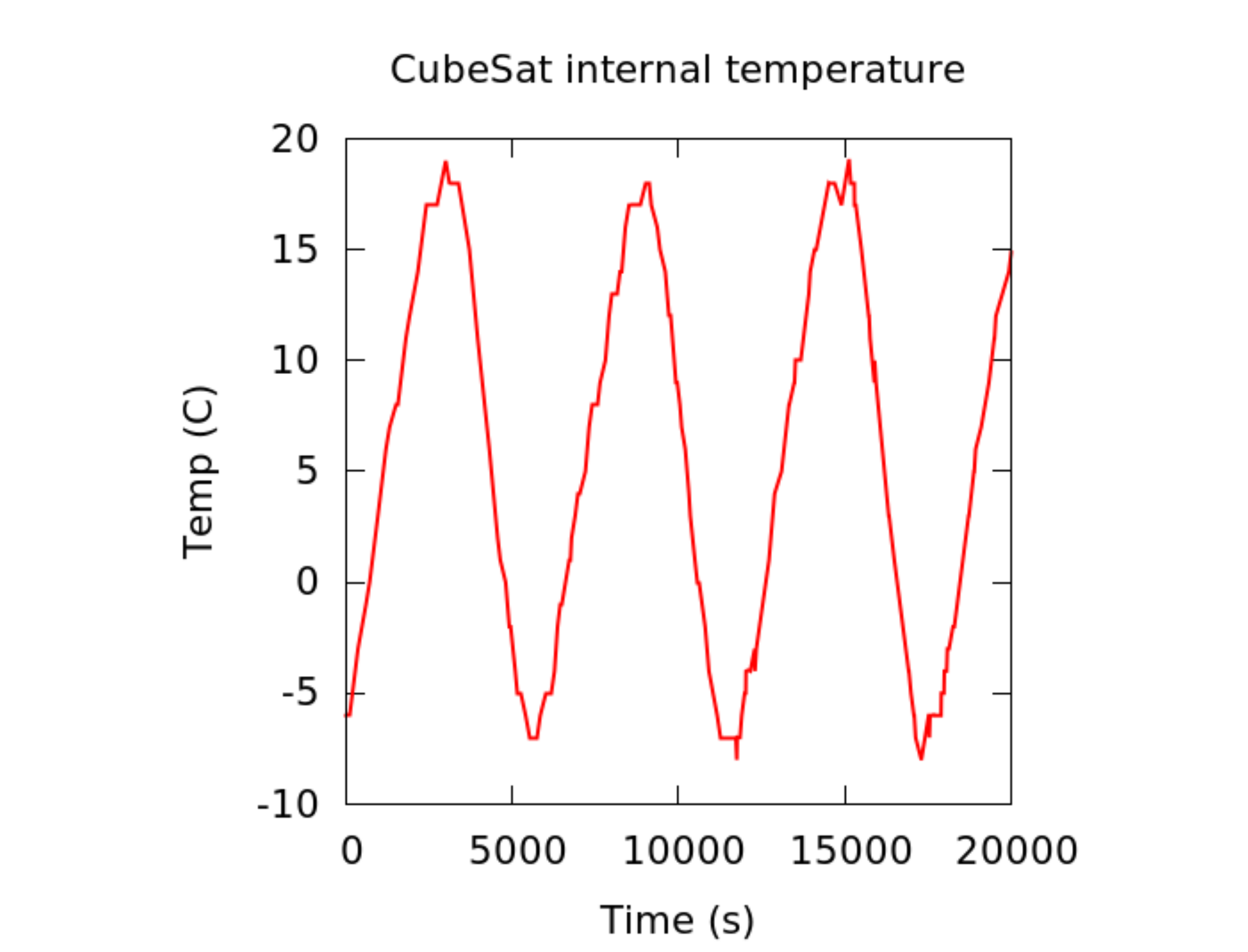}
\caption{\small {\textbf Figure 7.} Snapshot of the internal temperature of various sub-systems on the GomX-1 spacecraft that orbited at a \SI{600}{\km} polar orbit.}
\label{fig:gomx1}
\end{center}
\end{figure}

The orbits being analysed have also been flown a number of times by CubeSats. From these spacecraft the effect of orbit on internal spacecraft temperature has been mapped out a number of times \cite{apdrad}. In Fig. \ref{fig:gomx1} the internal temperature of the onboard computer from the GomX-1 spacecraft \cite{alminde14} varies between -\SI{7}{\celsius} and \SI{20}{\celsius}. The SPEQS thermal management system has been designed to deal with such a temperature variation.

\section{After SpooQy-Sat} \label{sec:outlook}
The rapid development of high performance capabilities of
nanosatellite systems will enable challenging and ambitious quantum
optics experiments in space. The SpooQy-Sat program is an example of how  cost effective small satellites can be used to demonstrate quantum technologies in space and to raise the technology readiness level of key components. Once the entangled photon sources are developed to a sufficiently mature stage it would be necessary to couple them to space optics capable of long distance transmission. 

High precision ADCS and pointing is an
area where concentrated effort has been applied due to various
applications ranging from high speed classical
communications~\cite{Kingsbury2014} to astronomical
observations~\cite{Koudelka2014}. Two-stage coarse and fine pointing
systems are being developed for several CubeSat
missions~\cite{Smith2011,Hemmati2013,Kingsbury2014}. 

A simultaneous transmission of an entangled photon pair to separate
ground stations has been a long term goal of quantum space
experiments~\cite{Ursin2009}. The requirement to point at two widely
separated targets necessitates large moving optical assemblies that
would be difficult to accommodate within a nanosatellite. Larger
satellites, such as microsats~\cite{Jennewein2012}, may be necessary
to meet SWaP requirements. However, the nano-microsat line is
beginning to blur with commercial services offering 12U CubeSat launch
and deployment into LEO for under USD1M. This expands the class of
missions that are feasible with small standardized containerized
platforms.

\section{Acknowledgments}
This work is supported by the National Research Foundation grant NRF-CRP12-2013-02 . J. A. Grieve is supported by the MOE grant MOE2012-T3-1-009. Valuable planning and design advise was provided by Alan Smith of Mullard Space Science Laboratory and Hans Kuiper of the aerospace department from TU Delft. 


\bibliographystyle{spiebib}   

\bibliography{report}   

\begin{thebibliography}{10}

\bibitem{ursin09}
{Ursin, R.}, {Jennewein, T.}, {Kofler, J.}, {Perdigues, J. M.}, {Cacciapuoti,
  L.}, {de Matos, C. J.}, {Aspelmeyer, M.}, {Valencia, A.}, {Scheidl, T.},
  {Acin, A.}, {Barbieri, C.}, {Bianco, G.}, {Brukner, C.}, {Capmany, J.},
  {Cova, S.}, {Giggenbach, D.}, {Leeb, W.}, {Hadfield, R. H.}, {Laflamme, R.},
  {Lütkenhaus, N.}, {Milburn, G.}, {Peev, M.}, {Ralph, T.}, {Rarity, J.},
  {Renner, R.}, {Samain, E.}, {Solomos, N.}, {Tittel, W.}, {Torres, J. P.},
  {Toyoshima, M.}, {Ortigosa-Blanch, A.}, {Pruneri, V.}, {Villoresi, P.},
  {Walmsley, I.}, {Weihs, G.}, {Weinfurter, H.}, {Zukowski, M.}, and
  {Zeilinger, A.}, ``{Space-QUEST, experiments with quantum entanglement in
  space},'' {\em Europhysics News}~{\bf 40}(3),  26--29 (2009).

\bibitem{ling12}
Ling, A. and Oi, D., ``{Small Photon-Entangling Quantum Systems (SPEQS) for LEO
  Satellites},'' {\em Proc. International Conference on Space Optical Systems
  and Applications (ICSOS) 2012, 11-1, Ajaccio, Corsica, France}  (October 9-12
  2012).

\bibitem{morong12}
Morong, W., Oi, D., and Ling, A., ``{Quantum Optics for Space Platforms},''
  {\em Optics and Photonics News}~{\bf 23}(10),  42--49 (2012).

\bibitem{scheidl13}
Scheidl, T., Wille, E., and Ursin, R., ``{Quantum optics experiments using the
  International Space Station: a proposal},'' {\em New J. Phys.}~{\bf 15},
  043008 (2013).

\bibitem{jennewein14}
Jennewein, T., Grant, C., Choi, E., Pugh, C., Holloway, C., Bourgoin, J.,
  Hakima, H., Higgins, B., and Zee, R., ``{The NanoQEY mission: ground to space
  quantum key and entanglement distribution using a nanosatellite},'' {\em
  Proc. SPIE}~{\bf 9254},  925402 (2014).

\bibitem{bourgoin13}
Bourgoin, J.-P., Meyer-Scott, E., Higgins, B.~L., Helou, B., Erven, C., Hübel,
  H., Kumar, B., Hudson, D., D'Souza, I., Girard, R., Laflamme, R., and
  Jennewein, T., ``{A comprehensive design and performance analysis of low
  Earth orbit satellite quantum communication},'' {\em New Journal of
  Physics}~{\bf 15}(2),  023006 (2013).

\bibitem{gerhardt11}
Gerhardt, I., Liu, Q., Lamas-Linares, A., Skaar, J., and Christian~Kurtsiefer,
  . V.~M., ``Full-field implementation of a perfect eavesdropper on a quantum
  cryptography system,'' {\em Nature Communications}~{\bf 2},  349 (2011).

\bibitem{opals14}
``{OPALS}.'' http://www.jpl.nasa.gov/news/news.php?feature=4402 (2014).
\newblock Accessed: 2015-06-30.

\bibitem{coopersmith11}
Coopersmith, J., ``{The cost of reaching orbit: Ground-based launch systems},''
  {\em Space Policy}~{\bf 27},  77--80 (May 2011).

\bibitem{woellert11}
Woellert, K., Ehrenfreund, P., Ricco, A.~J., and Hertzfeld, H., ``{Cubesats:
  Cost-effective science and technology platforms for emerging and developing
  nations},'' {\em Advances in Space Research}~{\bf 47},  663--684 (2011).

\bibitem{tan13}
Tan, Y.~C., Chandrasekara, R., Cheng, C., and Ling, A., ``Silicon avalanche
  photodiode operation and lifetime analysis for small satellites,'' {\em Opt.
  Express}~{\bf 21},  16946--16954 (Jul 2013).

\bibitem{tang_icsos_14}
Tang, Z., Chandrasekara, R., Sean, Y.~Y., Cheng, C., Wildfeuer, C., and Ling,
  A., ``{High Altitude Demonstration of Correlated Photon Source for
  Satellite-Based Quantum Key Distribution},'' {\em Proc. International
  Conference on Space Optical Systems and Applications (ICSOS) 2014, S7-4,
  Kobe, Japan, May 7-9 (2014)}  (2014).

\bibitem{tang14}
Tang, Z., Chandrasekara, R., Sean, Y.~Y., Cheng, C., Wildfeuer, C., and Ling,
  A., ``Near-space flight of a correlated photon system,'' {\em Scientific
  Reports}~{\bf 4},  6366 (2014).

\bibitem{tan15}
Tan, Y.~C., Chandrasekara, R., Cheng, C., and Ling, A., ``Radiation tolerance
  of opto-electronic components proposed for space-based quantum key
  distribution,'' {\em Journal of Modern Optics}  (2015).
\newblock DOI:10.1080/09500340.2015.1046519.

\bibitem{chandrasekara15_spie}
Chandrasekara, R., Tang, Z., Tan, Y.~C., Cheng, C., Wildfeuer, C., and Ling,
  A., ``Single photon counting for space based quantum experiments,'' {\em
  Proceedings of SPIE}~{\bf 9492},  949209 (2015).

\bibitem{cheng15}
Cheng, C., Chandrasekara, R., Tan, Y.~C., and Ling, A., ``Space qualified
  nanosatellite electronics platform for photon pair experiments,'' {\em
  arXiv:1505.06523}  (2015).

\bibitem{chandrasekara15_spie2}
Chandrasekara, R., Tang, Z., Tan, Y.~C., Cheng, C., Septriani, B., Durak, K.,
  Grieve, J., and Ling, A., ``{Deploying quantum light sources on
  nanosatellites I: lessons and perspectives on the optical system},'' {\em
  Proceedings of SPIE}~{\bf 9615},  961528 (August 2015).

\bibitem{gomx2}
``{GomX-2}.''
  \url{http://www.nasa.gov/mission_pages/station/research/experiments/1328.html}.
\newblock Accessed: 2015-06-30.

\bibitem{luo14}
Sha, L., Mouthaan, K., Seng, S.~W., Hiang, G.~C., and Ling, A., ``{Galassia
  System and Mission},'' {\em Small Satellite Conference} (SSC14-XI-2) (2014).

\bibitem{trl}
``{TRL}.''
  \url{https://www.nasa.gov/directorates/heo/scan/engineering/technology/txt_accordion1.html}.
\newblock Accessed: 2015-07-01.

\bibitem{swartwout14}
Swartwout, M., ``The first one hundred cubesats: A statistical look,'' {\em
  JoSS}~{\bf 2}(2),  213 (2014).

\bibitem{bouwmeester08}
Bouwmeester, J., Aalbers, G.~T., and Ubbels, W.~J., ``{Preliminary mission
  results and project evaluation of the delfi-c3 nano-satellite},'' tech. rep.,
  TU Delft (2008).

\bibitem{nanoracks}
``Nanoracks.'' http://nanoracks.com/.
\newblock Accessed: 2015-06-30.

\bibitem{spaceflight}
``{Space Flight Services}.'' http://www.spaceflightindustries.com/.
\newblock Accessed: 2015-06-30.

\bibitem{fitsat1}
``{FITSat-1}.''
  \url{https://directory.eoportal.org/web/eoportal/satellite-missions/f/fitsat-1}.
\newblock Accessed: 2015-06-30.

\bibitem{trojek08}
Trojek, P. and Weinfurter, H., ``{Collinear source of polarization-entangled
  photon pairs at nondegenerate wavelengths},'' {\em Applied Physics
  Letters}~{\bf 92},  211103 (2008).

\bibitem{qiao13}
Li~Qiao, C.~R. and Dempster, A.~G., ``Analysis and comparison of cubesat
  lifetime,'' {\em Proceedings of the 12th Australian Space Conference} ,
  249--246 (September 2013).

\bibitem{gomspace}
``{GOMSpace}.'' \url{http://gomspace.com/}.
\newblock Accessed: 2015-07-24.

\bibitem{nanoracksicd}
``{Nanoracks ICD v0.36}.''
  \url{http://nanoracks.com/wp-content/uploads/Current_edition_of_Interface_Document_for_CubeSat_Customers.pdf}.
\newblock Accessed: 2015-07-24.

\bibitem{apdrad}
Kataoka, J., Toizumi, T., Nakamori, T., Yatsu, Y., Tsubuku, Y., Kuramoto, Y.,
  Enomoto, T., Usui, R., Kawai, N., Ashida, H., Omagari, K., Fujihashi, K.,
  Inagawa, S., Miura, Y., Konda, Y., Miyashita, N., Matsunaga, S., Ishikawa,
  Y., Matsunaga, Y., and Kawabata, N., ``In-orbit performance of avalanche
  photodiode as radiation detector on board the picosatellite cute-1.7+apd
  ii,'' {\em Journal of Geophysical Research}~{\bf 115},  A05204 (2010).

\bibitem{alminde14}
Alminde, L.~K., Kaas, K., Bisgaard, M., Christiansen, J., and Gerhardt, D.,
  ``{GOMX-1 Flight Experience and Air Traffic Monitoring Results},'' {\em Small
  Satellite Conference}  (2014).

\bibitem{Kingsbury2014}
Kingsbury, R., Riesing, K., and Cahoy, K., ``Design of a free-space optical
  communication module for small satellites,'' in [{\em {28th Annual AIAA/USU
  Conference on Small Satellites}}{\nolinebreak\hspace{0.1em}]},  (2014).
\newblock SSC14-IX-6.

\bibitem{Koudelka2014}
Koudelka, O. and M.~Unterberger, P.~R., ``{Nanosatellites - the BRITE and
  OPS-SAT missions},'' {\em Elektrotechnik und Informationstechnik}~{\bf 131},
  178 (2014).

\bibitem{Smith2011}
Smitha, M.~W., Seagerb, S., Ponga, C.~M., Villase{\~n}orc, J.~S., Rickerc,
  G.~R., Millera, D.~W., Knappa, M.~E., Farmerb, G.~T., and Jensen-Clemd, R.,
  ``{ExoplanetSat: Detecting transiting exoplanets using a low-cost CubeSat
  platform},'' {\em Proceedings of SPIE}~{\bf 7731},  773127 (2010).
\newblock {Space Telescopes and Instrumentation 2010: Optical, Infrared, and
  Millimeter Wave}.

\bibitem{Hemmati2013}
Hemmati, H., ``{Laser-Communications with Lunar CubeSat},'' in [{\em {2nd
  International Workshop on LunarCubes}}{\nolinebreak\hspace{0.1em}]},  (2013).

\bibitem{Ursin2009}
Ursin, R., Jennewein, T., and Zeilinger, A., ``Space-quest: quantum physics and
  quantum communication in space,'' {\em {Proceedings of SPIE}}~{\bf
  7236}(723609) (2009).

\bibitem{Jennewein2012}
D'Souza, I., Hudson, D., Evans, C., Choi, E., Jennewein, T., and Sarda, K.,
  ``{The QEYSSat Mission: Demonstrating Global Quantum Key Distribution Using a
  Microsatellite},'' in [{\em {26th Annual AIAA/USU Conference on Small
  Satellites}}{\nolinebreak\hspace{0.1em}]},  (2012).
\newblock SSC12-IX-1.

\end{thebibliography}

\end{document}